\documentclass[12pt]{article}
\usepackage{graphicx}
\usepackage{hyperref}
\usepackage{amsmath}
\usepackage{amscd,amssymb}
\usepackage[cmtip,arrow,
matrix]{xy}

\usepackage{curves}

\newcommand{\cF}{\mathcal{F}}

\newcommand{\cG}{\mathcal{G}}

\def\cA{{\mathcal A}}

\def\cP{{\mathcal P}}
\def\cT{{\mathcal T}}

\def\cF{{\mathcal F}}

\def\cG{{\mathcal G}}

\def\cQ{{\mathcal Q}}

\def\mg{{\mathfrak g}}

\newcommand{\beq}{\begin{eqnarray}}
\newcommand{\eeq}{\end{eqnarray}}
\numberwithin{equation}{section}

\begin{document}


\begin{center}
{\large\bf D-Branes on Spaces Stratified Fibered Over Hyperbolic
Orbifolds}
\end{center}

\vspace{0.1in}

\begin{center}
{\large A. A. Bytsenko $^{(a)}$ \footnote{aabyts@gmail.com}, M.
Chaichian $^{(b)}$ \footnote{masud.chaichian@helsinki.fi} , M. E.
X. Guimar\~aes $^{(c)}$ \footnote{emilia@if.uff.br}

\vspace{5mm} $^{(a)}$ {\it Departamento de F\'{\i}sica,
Universidade Estadual de Londrina\\ Caixa Postal 6001,
Londrina-Paran\'a, Brazil}

\vspace{0.2cm} $^{(b)}$ {\it
Department of Physics, University of Helsinki\\
P.O. Box 64, FI-00014 Helsinki, Finland}

\vspace{0.2cm} $^{(c)}$ {\it Instituto de F\'{\i}sica,
Universidade Federal Fluminense \\ Av. Gal. Milton Tavares de
Souza, \\ s/n CEP 24210-346, Niter\'oi-RJ, Brazil}}
\end{center}

\vspace{0.1in}

\begin{abstract}
We apply the methods of homology and K-theory for branes wrapping
spaces stratified fibered over hyperbolic orbifolds. In addition,
we discuss the algebraic K-theory of any discrete co-compact Lie
group in terms of appropriate homology and Atiyah-Hirzebruch type
spectral sequence with its non-trivial lift to K-homology. We
emphasize the fact that the physical D-branes properties are
completely transparent within the mathematical framework of
K-theory. We derive criteria for D-brane stability in the case of
strongly virtually negatively curved groups. We show that branes
wrapping spaces stratified fibered over hyperbolic orbifolds carry
charge structure and change the additive structural properties in
K-homology.
\end{abstract}
\vfill



\newpage

\section{Introduction}
\label{Introduction}

In this paper we will elucidate the observation that methods of
algebraic K-theory and K-homology are appropriate tools in
classyfy branes wrapping spaces stratified fibered over hyperbolic
orbifolds. Generally speaking, D-branes are complicate objects and
cannot be considered as just subspace in an ambient spacetime.
Their dynamics require a more abstract mathematical tools (such as
of a derived category \cite{Douglas}, for example). Nevertheless
in algebraic K-theory and K-homology D-branes can be realized
naturally as an appropriate spacetime background with reveals
various properties of their quantum dynamics that could not be
otherwise detected if one classified the brane worldvolumes using
only singular homology \cite{Reis,Bytsenko}.

This paper is outlined as follows. In Sect. \ref{Hyper} we select
the requisite mathematical material used for investigation of
spaces stratified fibered over orbifolds. Our purpose there is to
discuss homology with {\it twisted} spectrum coefficients. The
main attention we pay for an Atiyah-Hirzebruch type spectral
sequence with it relation to (ordinary) homology.

Then we continue with algebraic K-theory of spaces stratified
fibered over hyperbolic orbifolds. Before discussing D-branes we
begin with some background material on stratified systems of
(abelian) groups over hyperbolic manifolds, and homology groups in
which the obstructions lie. We follow lines and conventions of
\cite{Quinn} which apply throghout. We examine homology spectrum
-- $\Omega$-spectrum, stratified system of abelian groups and
virtually negatively curved groups.

In the course of paper we follow the lines of \cite{Reis} and
assume the folowing definitions: A D-brane on background $\mathcal
B$ is a triple $[X,{F},\psi]$, where $X$ is a ${\it Spin}^{\mathbb
C}$ submanifold of $\mathcal B$, interpreted as a D-brane in
string theory, $F$ is the Chan-Paton boundle on $X$ and $\psi :
X\hookrightarrow {\mathcal B}$ is the embedding. A D-brane can be
interpreted as the homology class of K-cycle $[X, F, \psi]\in
K(\mathcal B)$;  we assume that the base space $\mathcal B$ is a
simply connected finite CW-complex (one can consider $\mathcal B$
to be only path-connected). There are many such situations in
which one is interested in the classification of D-branes in X. In
fact, many of the examples considered in literature fall into this
category. For instance, both the lens spaces $L(p;
q_1,\ldots,q_n)$ and the real projective spaces ${\mathbb
R}P^{2n+1}$ are circle bundles over ${\mathbb C}P^n$ \cite{Reis}.
Furthermore, the very application of vector bundle modification
identifies those D-branes such that one is a spherical fibration
over the other.

In Sect. \ref{hyperbolic} we analyze the additive structure in the
homology class of K-cycle and appropriate structural properties
for D-branes wrapping spaces stratified fibered over hyperbolic
orbifolds.

Finally in Sect. \ref{conclusions} we derive stability criteria
for D-branes wrapping spaces stratified fibered over hyperbolic
orbifolds trying to put physical notions in a mathematically
rigorous setting as much as possible.

\section{Branes on spaces stratified fibered over orbifolds}
\label{Hyper}

Let ${\overline X}$ be a closed (means compact and without
boundary) connected Riemannian manifold with strictly negative
sectional curvatures and let $\{X_j\}_{j =1}^\infty$,\, $ X_1
\subset X_2 \subset X_3 \subset ... $ be a sequence of connected
compact smooth manifolds. Let $F$ be a finite group which acting
on $\overline X$ via isometries and on each $X_{j}$ via smooth
maps. (The action of $F$ on $X_j$ need not to be free but on
$\overline X$ is free.) Assume also the smooth embedding
$X_{j}\subset X_{j+1}$ is both ${F}$-equivariant and
$j$-connected. Let $X_{\infty} = \bigcup_{j=1}^{\infty}X_{j}$ and
give $X$ the direct limit topology. The induced action of ${F}$ on
$X$ is free. Assuming that $\{M^j\}_{j=1}^{\infty}$ is the orbit
space ${\overline X}\times X_{j}$ under the diagonal action of
$F$, and
\begin{equation}
M^\infty = \lim_{\stackrel{\scriptstyle\longrightarrow}
{\scriptstyle j}}\, M^j\,.
\end{equation}
Let $X$ be the orbit space ${\overline X}/F$ and
$\{p_j\}_{j=1}^{\infty} :  {M}^{j}\rightarrow {X}$ be the map
induced from the canonical projection of ${\overline X}\times
X_{j}$ onto $\overline X$. The set $\{p_{j}\}$ is stratified
systems of fibrations on $X$. Let as before $\cG(Y)$ be the stable
topolgical pseudo-isotopy $\Omega$-spectrum associated to a
topological space $Y$; then we have to define a class of maps
$p_{\infty}$.

{\bf Stratified system of abelian groups.} With a stratified
system of fibrations $p: M \rightarrow {\mathcal B}$ one can
associate a homology $\Omega$-spectrum ${\mathbb H}({\mathcal B};
\cG(p))$ with tvisted spectrum coefficients and a map of $
\Omega$-spectra: $ {\mathbb H}({\mathcal B};
\cG(p))\longrightarrow \cG(M)$. A functor $\cG$ from spaces to
spectra is said to be homotopy invariant provided a homotopy
equivalence of spaces $X\rightarrow Y$ induces a homotopy
equivalence of spectra ${\cG}(X)\rightarrow {\cG}(Y)$ {\rm
\cite{Quinn}}. A homology $\Omega$-spectrum ${\mathbb H}({\mathcal
B}; \cG(p))$ is defined and discussed in the lectures
\cite{Bousfield}. One can construct Atiyah-Hirzebruch type
spectral sequence \cite{Quinn,Farrell}:

Suppose $\cG$ is a homotopy invariant spectrum valued functor of
spaces, and $p: M\rightarrow {{\mathcal B}}$ is an stratified
system of fibrations discussed above. Then the homotopy groups
$\pi_q\cG(p^{-1}(x))$ form stratified systems of groups over $X$,
denoted by $\pi_q\cG(p)$, and there is a (homological) spectral
sequence with
\begin{equation}
E_{p, q}^2 = {H}_p({\mathcal B}; \pi_q\cG(p)) \label{ss}
\end{equation}
which abuts to $\pi_{p+q}{\mathbb H}({\mathcal B};
\pi_q(\cG(p)))$, where $\pi_q\cG(p(x))$ denotes the stratified
system of abelian groups over ${\mathcal B}$ with the group
$\pi_q\cG(p^{-1}(x))$ corresponding to the point $x\in {\mathcal
B}$.

Let $\mathcal A$ be a stratified system of abelian groups. The
homology groups $H_j({\mathcal B}; {\mathcal A})$ can be
calculated as follows. A triangulation of a compact manifold $X$
consists of finite family of closed subsets $\{{\cT}_j\}_{j =1}^n$
that cover $X$, and family of homeomorphisms $\varphi_j :
{\cT}_j^{\prime}\rightarrow {\cT}_j$, where each
${\cT}_j^{\prime}$ is a Euclidean simplex. (In the case of compact
surface ${\mathcal B} = S$ each ${\cT}_j^{\prime}$ is a triangle
in the plane ${\mathbb R}^2$, i.e., a compact subset of ${\mathbb
R}^2$ bounded by three distinct strait lines.) Let $\cT$ be such a
triangulation that each strata ${\cT}_j$ is a subcomplex. Then a
stratified system $\cA$ restricted to each open simplex is a
constant system of coefficients and
\begin{equation}
H({\mathcal B}; \cA)=\bigoplus_j H_j(X; \cA)
\end{equation}
is the homology of finite chain complex
\begin{eqnarray}
C & := & C_0 \longrightarrow C_1 \longrightarrow ...
\longrightarrow C_n\,,
\nonumber \\
C_j & = & \bigoplus_{s \in {\cT}_j}\cA(\widehat{s})\,.
\end{eqnarray}
Here $\widehat{s}$ denotes the barycenter of $s$. In fact
$n$-dimensional $s$-simplex can be decomposed on $(n+1)!$ small
$n$-dimensional simplexes. Vertexes of new siplexes are the
centers of gravity of faces of initial simplex. Set $\{x_{j}\}$ of
centers is a set of vertexes of some simplex of barycentric
subdivision if the corresponding faces can be composed to chain of
consecutively enclosured faces.

{\bf The $E^2_{0,q}$-terms.} These terms in spectral sequence
(\ref{ss}) can be evaluated in the following manner. Let
${\mathfrak G}$ be the extension of $\pi_1({\overline X})$
determined by the action of $F$ on $\overline X$. In fact the
group $\mathfrak G$ is isomorphic to the factor group
$\pi_1(M)/\pi_1(X)$. Let $\cF ({\mathfrak G})$ be the category
with objects the finite subgroups of $\mathfrak G$. For each
element $\gamma \in {\mathfrak G}$ we can determine a morphism
$G_1\rightarrow G_2$ provided $\gamma G_1\gamma^{-1}\subseteq
G_2$. Then $E^2_{0, q}$-terms are isomorphic to the direct limit
\begin{equation}
E^2_{0, q} \,\,\stackrel{\rm
isomorphism}{-\!\!\!-\!\!\!-\!\!\!-\!\!\!-\!\!\!\longrightarrow}
\,\,\!\!\!\lim_{\stackrel{\longrightarrow} {\scriptstyle G\in
\cF({\mathfrak G})}} \!\!\pi_{q}(\cG(X/p (G)))\,,
\label{isomorphism}
\end{equation}
where $p: {\mathfrak G} \rightarrow {F}$ is the canonical
projection and $X/p (G)$ is the orbit space. This isomorphism can
be demonstrate by using of a basic result of E. Cartan :

{\bf (i)} First let us remind some conventions which will be
required. Let $A_{\mu}$ be the Yang-Mills connection associated
with some quantum field model and let $W$ be a (affine) space of
all connections $A_{\mu}$. The action of group $G$ on $A_{\mu}$ is
given by: $ A_{\mu}\rightarrow {g}A_{\mu}{g}^{-1} +
\partial_{\mu}{g}{g}^{-1}, \, {g}\in {G}, $ where the infinite
group ${G}$ is the group of gauge transformations. Set of classes
of equivalence of space $W$ with respect to ${G}$ action is called
{\it the orbit space}. Remind also that if geodesic $\gamma$ in
$X$ has ends belonging $Y\subset X$ then it entirely lie in $Y$;
manifold $Y$ satisfies this condition is called {\it completely
geodesic submanifold} in $X$.

{\bf (ii)} Let us consider a map $ \varphi\!: \,G/\Gamma
\rightarrow G,\,\, \varphi\!: \, g\Gamma\rightarrow
\alpha(g)g^{-1}, $ where $\alpha$-involutery authomorphism of
group $G$ satisfies: $\alpha^2= Id, \alpha \neq {Id},\,
(G_{\alpha})_0\subset \Gamma\subset G_{\alpha}$. Here $G_{\alpha}$
is a set of fixed points of the authomorphism $\alpha$, and
$(G_{\alpha})_0$ is a component of unit of $G_{\alpha}$. First
note that $\varphi$ is the diffeomorphism $G/\Gamma$ on closed
completely geodesic manifold $ X_{\alpha}\!: X_{\alpha} = \{g \in
G|\alpha(g)g={Id}\}\,. $ A differential $\alpha$ in point ${Id}$,
$d\alpha_{Id}$, is the involutery authomorphism of Lie algebra $A$
of group $G$. The operator $d\alpha_{Id}$ decomposes the space $A$
on two subspaces $P$ and $Q$ associated with eigenvalues $\pm 1$
of $ d\alpha_{Id}: P = \{\cP \in {A}|d\alpha_{Id}(\cP)=\cP\},\, Q
= \{\cQ \in {A}| d\alpha_{Id}(\cQ)= -\cQ\}. $

It can be shown that the algebra $Q$ of Lie group $\Gamma$ is a
$P$-orthogonal (with respect to scalar product on $A$) to
$Q$-subspace in $A$. The subspace $P$ is invariant with respect to
adjoint representation $Ad(\Gamma)$ (see for detail
\cite{Helgason}). The space $X_{\alpha}$ is the closed submanifold
in $G$, and ${\rm dim}\, X_{\alpha} = {\dim} (G/\Gamma)$.
\footnote{ Indeed, the space $X_{\alpha}$ is a kernel of the map
$\varphi$. In order to find ${\dim}\, X_{\alpha}$ it is convenient
to calculate the kernel of tangent map $ T_{g}\varphi:
T_{g}G\rightarrow T_{\alpha(g)g}G $ in a point $g$. ${\rm Ker}\,
T_{g}\varphi = {g}W_{g}$, where $W \equiv W_{g} = \{\cP\in
{A}|d\alpha_{Id}(\cP) + {\rm Ad}(g)\cP = 0\}$. It means that ${\rm
Ker}\, T_{g}\psi = P$, ${\rm rank} \,\varphi = {\dim}\,{A} - {\rm
dim}\,{\rm Ker}\,\varphi = {\dim}\, Q$, and it is not depend on
choice of a poit $g$. It follows that $X_{\alpha}$ is the close
submanifold in $G$ and ${\rm dim}\, X_{\alpha}= {\rm dim}
\,(G/\Gamma)$. } Also $X_{\alpha}$ is the completely geodesic
submanifold in $G$. It is known that geodesics (with respect to
invariant metric) in space $G/\Gamma$ are orbits of one-parameter
subgroups of group $G$ \cite{Helgason}. Let $\gamma$ be a geodesic
in a group $G$ tangent to $X_{\alpha}$ in a point $g$: $\gamma_t =
g{\bf e}[t\cP]$, where ${\bf e}[-]\equiv \exp(-)$, $\cP \in A$,
and $g\cdot \cP \in T_{g}X_{\alpha} = {\rm Ker}\, T_{g}\varphi =
{g}W{g}$. We have
\begin{eqnarray}
&& \alpha(\gamma_t)\gamma_t  =  \alpha(g{\bf e}[t\cP])g{\bf
e}[t\cP] = \alpha(g){\bf e}[td\alpha_{Id}(\cP)]g{\bf e}[t\cP]
\nonumber \\\
&& \stackrel {{\rm Ker} T_{g}\varphi  =  g Wg}
{=\!\!=\!\!=\!\!=\!\!=\!\!=\!\!=\!\!\Longrightarrow} \alpha(g){\bf
e}[td\alpha_{Id})(\cP)]\mg{\bf e}[t\cP]  = \alpha(g){\bf e}[{\rm
Ad}(g)t\cP]g{\bf e}[t\cP]
\nonumber \\\
&& \Longrightarrow \alpha(\gamma_t)\gamma_t  =  \alpha(g)g{\bf
e}[-t\cP]g^{-1}g {\bf e}[t\cP]  = \alpha(g)g = {Id}\,.
\end{eqnarray}
Let us proof that $X_{\alpha}= {\rm Im}\,\varphi$. We have
$X_{\alpha} \supset {\rm Im}\,\varphi$; it can be prooved that
$X_{\alpha} \subset {\rm Im}\,\varphi$. In $X_{\alpha}$ there is a
finite number $N$ of points $\{g_j\}$ and geodesics $\{\gamma_j\}$
which joins $g_{j-1}$ with $g_j$, such that $g_N \in {\rm
Im}\,\varphi$. Let $g_{N-1}= g_N{\bf e}[t\cP]$, $\cP\in W_{g_N}$
for a some number of $t$. Then we have:
\begin{eqnarray}
&& \alpha({\bf e}[-(1/2)t\cP]g_N)({\bf e}[-(1/2)t\cP]g_N)^{-1} =
{\bf e}[-(1/2)td\alpha_{Id}(\cP)]\alpha(g_N)g_N^{-1} {\bf
e}[(1/2)t\cP]
\nonumber \\\
&& = {\bf e}[(1/2)t{\rm Ad}(g_N)\cP]g_N{\bf e}[(1/2)t\cP] =
g_N{\bf e}[t\cP] = g_{N-1}\,.
\end{eqnarray}

{\bf (iii)} Finally repeating this procedure for $g_j,  0\leq
j\leq N-1$ we get $g_0\in {\rm Im} \,\varphi$. Due to E. Cartan
any orbit of a group $G$ can be imbeded in $G$ as a completely
geodesic submanifold. For any compact subgroup $\Gamma\subset G$
exists involutory authomorphism $\alpha$, leaving subgroup
$\Gamma$ fixed, it follows that all orbits of group $G$ can be
realized as completely geodesic manifolds.


\section{Strongly virtually negatively curved groups}
\label{hyperbolic}

\label{Hyper2} Let $X \cG({\mathcal B}; p)$ be a cofiber (in
category of spectra) which in fact is also an $\Omega$-spectrum.
In the case when $X$ is a contractible the following result holds
\cite{Farrell}:
\begin{eqnarray}
Wh_n({G_M})\otimes {\mathbb Q} & \cong & \bigoplus_{j=0} H_{j}(X;
Wh_{n-j}(G_y)\otimes {\mathbb Q})\,,
\\
K_n({\mathbb Z}{G_M})\otimes {\mathbb Q} & \cong & \bigoplus_{j=0}
H_{j}(X; K_{n-j}({\mathbb Z}G_y)\otimes {\mathbb Q})\,,
\end{eqnarray}
where $\pi_1(M)$ has been denoted by $G_M$, $G_y =
\pi_1(p^{-1}(y))$ for $y\in {\mathcal B}$, $Wh_n$ is the Whitehead
group and $Wh_{n-\ell}(G_y)\otimes {\mathbb Q}$,
$K_{n-\ell}({\mathbb Z}G_y)\otimes {\mathbb Q}$ are the suitable
stratified systems of abelian groups over ${\mathcal B}$. By
definition a group $G_M$ is {\it strongly virtually negatively
curved} if $G_M$ is isomorphic to $\pi_1 (M)$, where $M$ is
constructed as before with constraint that $X$ is contractible.
Note that any discrete co-compact subgroup $\Gamma$ of a Lie group
$G$, where $G$ is either $O(n, 1), \\ U(n, 1), Sp(n, 1)$, or
${F_4}$, is strongly virtually negatively curved.

Let ${{\mathcal B}} = G/{\Gamma}$ be an irreducible rank one
symmetric space and ${\Gamma}\subset G$ a maximal compact
subgroup. The following statement holds \cite{Farrell}: The
algebraic K-theory of any discrete co-compact subgroup of a Lie
group $G$, where $G= O(n,1),\, U(n,1),\, Sp(n,1),\, F_4$ can be
calculate in terms of the homology of the double coset space
${G_M}\backslash {{\mathcal B}} \equiv {G_M}\backslash
G/{\Gamma}$. As a consequence the following result follows:
\begin{equation}
K_n({\mathbb Z}{G_M})\otimes {\mathbb Q} \cong  \bigoplus_{j=0}
H_j({G_M}\backslash G/{\Gamma};  p_{n-j})\,, \label{Hom}
\end{equation}
where $p_n$ is a stratified system of $\mathbb Q$ vector space
over ${G_M} \backslash {G}/{\Gamma}$ and the vector space
$p_n({G_M} \mg {\Gamma})$ corresponding to the double coset ${G_M}
\mg {\Gamma}$ is isomorphic to $K_\sharp({\mathbb Z}({G_M} \cap
{\mg} {\Gamma}{\mg}^{-1}))\otimes {\mathbb Q}$. In addition ${G_M}
\cap {\mg}{\Gamma}{\mg}^{-1}$ is a finite subgroup of $G_M$. Any
discrete co-compact subgroup of Lie group $G$, where $G$ is either
$O(n,1), U(n,1), Sp(n,1)$ or $F_4$, is strongly virtually
negatively curved. Thus we obtain the additive structure in the
homology class of K-cycle and appropriate structural properties
for D-branes wrapping spaces stratified fibered over hyperbolic
orbifolds.

\section{Conclusions}
\label{conclusions}

The subject of this paper the intriguing relationship between
D-branes on spaces stratified fibered over negatively curved
orbifolds and algebraic K-theory. The mathematical methods are
based in Atiyah-Hirzebruch type spectral sequence which we used
for stratified system of (abelian) groups. By the use of this
technique we obtain results in the structure of the K-homology of
strongly virtually negatively groups. In summary we present
discussion and main results obtained in this paper:

{\bf (i)}\, A D-brane wrapping homologically nontrivial cycle can
nevertheless be unstable, if for some $X_{j+k}\subset X$ the
following condition holds \cite{Maldacena}: $ PD(X_j \subset
X_{j+k}) = \omega_3(X_{j+k}) + [H]\vert_{X_{j+k}}\,. \label{COND}
$ In this equation the left hand side denotes the Poincar\'{e}
dual of $X_j$ in $X_{j+k}$. One can use a mathematical algorithm,
the Atiyah-Hirzebruch spectral sequence (see Sect. \ref{Hyper}),
to determine which homology classes lift to K-theory classes (that
is, to determine which D-branes are stable and which are not
allowed) -- it gives the rigorous stability criterion
\cite{Reis}.

Suppose stratified system of fibrations on $X$ extends through the
spectral sequence as a non-trivial element of homology groups
$E_{p,q}^2$. The following result is effective for the rational
calculation of many K-groups (cf. Theorem 2 of \cite{Farrell}): If
as before $p: M\rightarrow {\mathcal B}$ is a stratified system of
fibrations and $X$ is aspherical with $Wh_n(\pi_1X\times
S^1)\otimes Q = 0$ for all integers $n$, then $X\cG({\mathcal B};
p)\otimes Q = 0$. The stratified system of fibrations could be
extends through the spectral sequence as a non-trivial element of
homology groups but it can have trivial lift to K-homology and it
vanishes in $E_{p,q}^\infty$.

{\bf (ii)}\, For each of the stratified systems of fibrations: $p:
M\rightarrow {\mathcal B}$ there is homotopy equivalence of
${\Omega}$-spectra: ${\cG}(M) \cong {\mathbb H}({\mathcal B};
\cG(p))\times X\cG({\mathcal B}; p). $ This equation is in
conformity with an Atiyah-Hirzebruch type stectral sequence
(\ref{ss}).
The Atiyah-Hirzebruch spectral sequence keeps track of the
possible obstructions for a homology cycle of starting from
${H}_p({\mathcal B}; \pi_q(\cG(p)))$ in (\ref{ss}) (the initial
terms are given in (\ref{isomorphism})), to survive to
$E_{p,q}^\infty$. For a number of explicit examples, strongly
virtually negatively curved $G_M$, the groups $G_y =
\pi_1(p^{-1}(y))$ are finite because they are isomorphic to
subgroups of $\Gamma$. As a consequence one can use the extensive
knowledge of the algebraic K-theory of finite groups
\cite{Quillen,Borel}. \footnote{ For example we have the following
result. If $G_M$ is a strongly virtually negatively curved then
$K_n({\mathbb Z}{G_M}) = 0$, $\forall \, n < - 1$. In the case
$n=-1$, $ K_{-1}({\mathbb Z}{G_M})\simeq
\lim_{{\stackrel{\longrightarrow} {\scriptstyle G\in
\cF({G_M})}}}K_{-1}({\mathbb Z}{G_M}), $ $K_{-1}({\mathbb
Z}{G_M})$ is a finitely generated abelian group; it is generated
by the images of $K_{-1}({\mathbb Z}{G_M})$, as $G$ varies over
the finite subgroups of $G_M$, under the map functorially induced
by the inclusion of $G$ into $G_M$. (Perhaps this statement is
true for a much larger class of groups $G_M$. In particular, in
\cite{Farrell} it has been conjectured that it is true for all
finitely generated subgroups $G_M$ of $GL_n({\mathbb C})$). }

{\bf (iii)}\, The homotopy groups $\pi_k{\cG}(p^{-1})$ form
stratified systems of groups $\pi_k{\cG}(p)$ over spaces ${X}$.
D-branes wrapping spaces stratified fibered over hyperbolic
orbifolds carry charge structure and change the additive
structural properties in K-homology.

\subsection*{Acknowledgments}

AAB and MEXG would like to acknowledge the Conselho Nacional de
Desenvolvimento Cient\'ifico e Tecnol\'ogico (CNPq, Brazil) and
Funda\c cao Araucaria (Parana, Brazil) for financial support. The
support of the Academy of Finland under the Projects No. 136539
and 140886 is gratefully acknowledged.

\end{document}